\documentclass[a4paper, 12pt, longbibliography]{article}

\newcommand{\iTitle}[1]{\begin{center}\Large\bf #1\end{center}}
\newcommand{\iAuthor}[1]{\begin{center}\small #1\end{center}}
\newcommand{\iAddress}[1]{\begin{center}\small\it #1\end{center}}
\newcommand{\iAbstract}[1]{\noindent {\large\bf Abstract}\\#1}

\usepackage{multirow}
\usepackage{geometry}
\usepackage{indentfirst} 
\usepackage[pagewise,switch]{lineno}
\usepackage{amsmath,amsfonts,amssymb}
\usepackage{graphicx}
\usepackage{subfigure}
\usepackage{color}
\usepackage{xcolor}
\definecolor{red}{rgb}{1,0,0}
\usepackage[colorlinks=true, linkcolor=blue, urlcolor=magenta, citecolor=red]{hyperref}
\newcommand{\refb}[1]{(\ref{#1})}

\usepackage{cite}

\newcommand{\bw}{\begin{widetext}}
\newcommand{\ew}{\end{widetext}}
\newcommand{\be}{\begin{equation}}
\newcommand{\en}{\end{equation}}
\newcommand{\bee}{\begin{equation}}
\newcommand{\ene}{\end{equation}}
\newcommand{\bea}{\begin{eqnarray}}
\newcommand{\ena}{\end{eqnarray}}
\newcommand{\eq}[1]{Eq.~(\ref{#1})}

\def\pslash{p\!\!\!\slash }

\def\to{\rightarrow}

\def\pslash{p\!\!\!\slash }

\def\ie{{\it i.e.}}

\begin{document}

\iTitle{Probing Noncommutativities of Phase Space by Using Persistent Charged Current and Its Asymmetry}

\vspace{0.3cm}
\iAuthor{Kai Ma\footnote{\href{mailto:makainca@yeah.net}{makainca@yeah.net}}, Ya-Jie Ren, Ya-Hui Wang}
\vspace{-0.6cm}
\iAddress{Department of Physics, Shaanxi University  of Technology, Hanzhong, 723001, Peoples Republic of China}

\vspace{0.5cm}
\noindent\rule[0.25\baselineskip]{\textwidth}{0.8pt}
\iAbstract{We study the physical properties of the persistent charged current in a metal ring on a noncommutative phase space, and temperature dependence of the noncommutative corrections are also analyzed. We find that the coordinate noncommutativity only affects the total magnitude of the current, and it is difficult to observe it. In contrast, the momentum noncommutativity can violate symmetry property of the current, and this violation of symmetry holds at a finite temperature. Based on this violation effect, we introduce an asymmetry observable to measure the momentum noncommutativity.
}

\vspace{0.3cm}

%\iKeywords{Noncommutative geometry, Aharonov-Bohm effect, Geometry phase}

\noindent\rule[0.25\baselineskip]{\textwidth}{0.8pt}

\tableofcontents
\newpage

\section{Introduction}\label{intro} 
One of the most important ingredients of Quantum Mechanics (QM) is the noncommutativity of physical quantities. Evolution of the quantum system is
governed by the elementary noncommutative algebras. The commutators  
among position and momentum operators, 
$[x_i,\, x_j]=0$, 
$[p_i,\, p_j]=0$, 
$[x_i,\, p_j]=i\hbar\delta_{ij}$, 
are fundamental algebras, and go into any quantum phenomena. However, there is no any guarantee
which can protect above commutators. Noncommutative space, which is
characterized by following algebra,
\bee\label{eq:ncdefine}
[x_{\mu}, x_{\nu}] = i \theta_{\mu\nu}\,,
\ene
is an extension of those relations, and parameterized by the totally anti-symmetric constant tensor $\theta_{\mu\nu}$ which has dimension of length-squared.
Mathematically, the noncommutative space can be a candidate of the physical background space-time. On the other hand, noncommutative property can also appear in new physics models, for instance, string theory embedded in a background magnetic field~\cite{Seiberg:1999vs}, quantum gravity~\cite{Freidel:2005me, Moffat:2000gr, Moffat:2000fv, Faizal:2013ioa}, and in real quantum system~\cite{Gamboa:2000yq, Ho:2001aa}. Due to that above extension
is fundamental, all the quantum physics are deformed, and scaled by the parameter 
$\theta_{\mu\nu}$. For instance, breaking of rotational symmetry~\cite{Douglas:2001ba, Szabo:2001kg}, distortion of energy levels of the atoms~\cite{Chaichian:2000si, Zhang:2004yu}, corrections on the spin-orbital interactions~\cite{Ma:2011gc,Basu:2012td,Deriglazov:2016mhk} as well as quantum speed of relativistic charged particles~\cite{Wang:2017azq,Wang:2017arq,Deriglazov:2015zta,Deriglazov:2015wde} and fluid dynamics~\cite{Das:2016hmc}.

%$\vec{\theta}\cdot\vec{B}=$
Most importantly, the spatial component of the noncommutative parameter ($\theta_{\mu\nu}$), $\theta^{ij}$, behaves like a permanent magnetic dipole moment, and gives a contribution $\epsilon_{ijk}\theta^{ij}B^{k}$ to the energy of system. This correction can affect the electromagnetic dynamics of charged particles. It has been shown in Refs.~\cite{Chaichian:2001pw, Chaichian:2000hy, Chaichian:2008gf, Ma:2016rhk, Anacleto:2016ukc, Mirza:2003hn, Li:2006pi, Mirza:2006ce, Li:2008zzq, Ma:2014tua} that magnetic flux due to the permanent dipole moment can modify the Aharonov-Bohm effect~\cite{Aharonov:1959fk} as well as the Aharonov-Casher effect~\cite{Aharonov:1984xb}. Furthermore, the magnetic flux can also generate persistent charged current in a metal ring. In Ref.~\cite{Liang:2014ija}, it was shown that the persistent charged current~\cite{PhysRevB.37.6050,Bleszynski-Jayich272,PhysRevB.81.155448} is sensitive to the spatial noncommtuativity, and can be employed to probe the low energy residual effects of possible new physics at high energy scale. However, the analysis in Ref.~\cite{Liang:2014ija} was done by using simple shift method which in general can not lead to gauge invariant result~\cite{Chaichian:2008gf, Ma:2014tua, Bertolami:2015yga, Ma:2016rhk}. Furthermore, the persistent current in real metal ring is also affected by temperature, cross section {\it etc}. Therefore, a more realistic investigation is necessary. In this paper, we will use the Seiberg-Witten (SW) map \cite{Seiberg:1999vs} which preserves the gauge symmetry to study the noncommutative effects on the persistent current in a metal ring, and analyze temperature dependence of the noncommutative corrections.

The contents of this paper are organized as follows: in Sec.~\ref{sec:hamiltonian} we introduce the  Hamiltonian of a charged particle on noncommutative phase space; in Sec.~\ref{sec:ncexpAB} we will study the noncommutative corrections on the persistent current in a metal ring, and based on  symmetry property of the persistent current we introduce a new asymmetry observable for probing the noncommutativity of phase space; in Sec.~\ref{sec:temperature}, we study the temperature dependence of the persistent current, as well as thermal effects on the asymmetry observable proposed in Sec.~\ref{sec:ncexpAB}; our conclusions are given in the final section, Sec. \ref{conclusion}.

\section{Noncommutative Hamiltonian}\label{sec:hamiltonian}
Noncommutative extension of the underlying space-time is characterized by the algebra
in \eqref{eq:ncdefine}. However, in consideration of that the momentum operators are 
defined as the derivatives with respect to coordinates, it is natural to consider 
further extension with noncommutative momentum operators defined by following algebra~\cite{Langmann:2002cc, Langmann:2003if, Li:2006cv, Li:2006pi, Muthukumar:2006ab, Liang:2014ija, Bertolami:2015yga, Ma:2016vac},
\bee\label{eq:npdefine}
[p_{\mu}, p_{\nu}] = i \xi_{\mu\nu}\,,
\ene
where $\xi_{\mu\nu}$ is also a totally anti-symmetric constant tensor with dimension
of squared energy. Physical effects of the above extension have been studied in various aspects~\cite{Langmann:2002cc, Langmann:2003if, Li:2006cv, Li:2006pi, Muthukumar:2006ab, Liang:2014ija,Bertolami:2015yga, Ma:2016vac}. And a more general extension on the generators of whole Poincare group was conducted recently in Ref.~\cite{Meljanac:2016jwk}. Therefore there is strong phenomenological motivations to study the extensions \eqref{eq:ncdefine} and \eqref{eq:npdefine} simultaneouslly.  

The two parameters $\theta_{\mu\nu}$ and $\xi_{\mu\nu}$ parameterize
the electromagnetic interactions on noncommutative space, and scale any new physics
due to non-triviality of the background space-time. In general, all their components 
can be nonzero. However, non-vanishing terms $\theta^{0i}$ and/or $\xi^{0j}$ can violate 
unitarity, therefore, in this paper, we consider only the spatial noncommutativity of 
both coordinates and momenta. While the coordinate noncommutativity can not be taken into
account by using simple shift method, following transformation can be used to realize
the algebra in \eqref{eq:npdefine}~\cite{Bertolami:2015yga},
\bee\label{eq:pshift}
p_{\mu} \to p_{\mu} - \frac{1}{2\hbar}\xi_{\mu\nu}x^{\nu}\,,
\ene
and can give gauge invariant physical contributions~\cite{Bertolami:2015yga,Ma:2016vac}. 
According to this transformation, the Lagrangian for a charged particle interacting with 
external electromagnetic fields can be written as~\cite{Ma:2016vac}
\bee\label{eq:lagrangian}
%\tilde{\mathcal{L}} =  
\mathcal{L} =
\bar{\psi}(x)( \pslash - Q ~\slash{\!\!\!\!A}_{NC} - m ) \psi(x) \,,
\ene
where $Q$ is the charge of matter particle in unite of $|e|$, and the total potential 
$A_{NC;\mu}$ is a sum of original one $A_{\mu}$ and an effective term $A_{\xi;\mu}$ 
emerging from the transformation \refb{eq:pshift},
\bee\label{eq:Ashift}
A_{NC;\mu} = A_{\mu} + A_{\xi;\mu}\,,\,\,\;
A_{\xi;\mu} = \frac{\xi }{2\hbar Q} \big(0, y,\, -x,\, 0  \big)\,.
%A_{\bar\theta;\mu} = \frac{\bar\theta c}{2\alpha^2\hbar e} \big(0, -y,\, x,\, 0  \big)\,.
\ene
In the above derivation of the explicit expression of $A_{\xi;\mu}$, we have defined 
the $\hat{z}$-axis along the direction of the vector $\vec{\xi}$, where the components of $\vec{\xi}$ are related to $\xi^{ij}$ by the relation $\xi_{i}=\epsilon_{ijk}\xi^{ij}/2$. In this configuration, the non-zero components are only $\xi^{12}=-\xi^{21}=\xi$, and the effect of momentum noncommutativity is an addition of a constant magnetic background field
\bee
\vec{B}_{\xi} 
= \vec{\nabla} \times \vec{A}_{\xi} 
= \big(0, 0, B_{\xi}  \big)
\,,~~~B_{\xi} = \frac{\xi }{\hbar Q} 
\ene
over the whole space. 

For the coordinate noncommutativity, it has been shown that the Seiberg-Witten 
(SW) map \cite{Seiberg:1999vs} from the noncommmutative space to the ordinary one preserves the 
gauge symmetry, and has been proved to be very useful to investigate various problems on 
noncommutative space~\cite{Ma:2014tua, Martin:2012aw, Brace:2001fj, Barnich:2001mc, Picariello:2001mu, Banerjee:2001un, Wang:2013iga, Wang:2012ye, Wang:2011ei}. Therefore we will
use the SW map to study the noncommutative corrections induced by the nontrivial algebra 
in \refb{eq:ncdefine}. For the $U(1)$ gauge symmetry the SW map is given by
\bea
\psi &\to & \psi - \frac{1}{2} Q \theta^{\alpha\beta} A_{NC;\alpha} \partial_{\beta}\psi\,,  \\
A_{NC;\mu} &\to & A_{NC;\mu} - \frac{1}{2} Q\theta^{\alpha\beta} A_{NC;\alpha} ( \partial_{\beta} A_{NC;\mu} + F_{NC;\beta\mu} )\,,
\ena
where $F_{NC;\mu\nu} = \partial_{\mu} A_{NC;\nu} -  \partial_{\nu} A_{NC;\mu} $ is the effective electromagnetic field strength tensor. Then in terms of the ordinary fields the noncommutative Lagrangian \refb{eq:lagrangian} can be written as,
\bee\label{ac-nc}
\mathcal{L}_{NC} =
\big( 1 - \frac{1}{4}Q\theta^{\alpha\beta} F_{NC;\alpha\beta} \big) \bar{\psi}(x) ( i\gamma_{\mu} D^{\mu}_{NC}  - m )  \psi(x) +
\frac{i}{2} Q\theta^{\alpha\beta} \bar{\psi}(x) \gamma^{\mu} F_{NC;\mu\alpha} D_{NC;\beta} \psi(x) \,,
\ene
where $D_{NC;\beta} =  \partial_{\beta} + i Q A_{NC;\beta}$ is the covariant derivative. This Lagrangian is gauge invariant because the noncommutative corrections depend only on the covariant derivative $D_{NC;\beta}$ and the electromagnetic field strength $F_{NC;\mu\nu}$. Therefore, the noncommutative corrections on the AB phase shift can be defined unambiguously, and can be interpreted consistently on the commutative and noncommutative phase spaces. The equation of motion can be obtained directly from this Lagrangian as follows,
\bee\label{ncmotioneq}
( i\gamma_{\mu} D^{\mu}_{NC}  - m )  \psi(x) +
\frac{i}{2} Q\big( 1 - \frac{1}{4}Q \theta^{\alpha\beta} F_{NC;\alpha\beta} \big) ^{-1} \theta^{\alpha\beta}\gamma^{\mu} F_{NC;\mu\alpha} D_{NC;\beta} \psi(x) =0\,,
\ene
where we have multiplied a factor $( 1 - Q \theta^{\alpha\beta} F_{NC;\alpha\beta}/4 )^{-1}$ to simplify the expression. By expanding $( 1 - Q \theta^{\alpha\beta} F_{NC;\alpha\beta}/4 ) ^{-1}$ with respect to the noncommutative parameter $\theta^{\mu\nu}$, one can see that the correction proportional to $Q\theta^{\alpha\beta} F_{NC;\alpha\beta}$ can be neglected at the first oder of the noncommutative parameter $\theta^{\mu\nu}$. In some studies this term was taken into account through the renormalization of the particle charge. However, for consistence, from here and after we will keep only the leading order terms. Therefore we will consider only the correction which involves the covariant derivative. Under this approximation, the equation of motion can be written as
\bea\label{ncacfinald}
&& ( i\gamma_{\mu} \mathcal{D}_{NC}^{\mu}  - m ) \psi(x) = 0 \,, 
\\[2mm]
\label{ncacfinald2}
&& \mathcal{D}_{NC}^{\mu}  = \big( \eta^{\mu}_{~~\beta} + \frac{1}{2} Q F_{NC}^{\mu\alpha}\theta_{\alpha\beta} \big) D^{\beta}_{NC} 
\equiv g^{\alpha}_{~~\beta}D^{\beta}_{NC} \,.
\ena
The deformed Dirac equation is similar to the Dirac equation in curved space-time with metric $g_{\alpha\beta}$. The leading order effect of the noncommutative phase space behaves like a perturbation on the flat space-time defined by the metric $\eta_{\alpha\beta}$. The perturbation depends on the external magnetic field, noncommutative parameters as well as their relative angles. The Hamiltonian resulting in equation of motion \eqref{ncacfinald} is
\begin{equation}
H_{NC} 
=
\beta m + g_{ij}\alpha^{i}(p^{j}-QA_{NC}^{j}) + Qg_{00}V_{NC}\,.
\end{equation}
The non-relativistic approximation can be obtained by using the well-know Foldy-Wouthuysen unitary transformation (FWUF)~\cite{Foldy:1949wa}, which approximately block diagonalizes the Dirac Hamiltonian by separating the positive and negative energy part of its spectrum. Neglecting the spin degree of freedom
the non-relativistic Hamiltonian is
\begin{equation}\label{eq:nonRelNCH}
H_{NC} = \frac{1}{2m_{\theta}}(\vec{p} - Q\vec{A}_{NC})^2\,,
\end{equation}
where
\begin{equation}\label{eq:massren}
m_{\theta} = \dfrac{m}{1-Q\dfrac{\vec{\theta}\cdot\vec{B}_{NC} }{2 \phi_{0} } }
\end{equation}
is the effective mass on the noncommutative phase space,
and $\phi_{0}=h/e$ is the fundamental magnetic flux. Because of that $|\theta||\xi| \ll \hbar^2$, and furthermore usually the external magnetic field $|\vec{B}|$ is much stronger then the noncommutative background $|\vec{B}_{\xi}|$, the mass renormalization effect given in \eqref{eq:massren} can be approximated as $m_{\theta} \approx m [1- Q \vec{\theta}\cdot\vec{B}/(2\phi_{0}) ]^{-1}$. This means that, in non-relativistic limit, coordinate noncommutativity only affect the effective mass of charged particles.

%%%%%%%%%%%%%%%%%%%
\section{Persistent Current and Asymmetry}\label{sec:ncexpAB}
The noncommutativity of space-time can distort the electromagnetic dynamical behavior of charge carriers, and their properties are essentially determined by the Hamiltonian given in \eqref{eq:nonRelNCH}. In this section we study the noncommutative effects on the persistent charged current in a metal ring~\cite{Liang:2014ija}. 
We will ignore the effects due to finite cross section of the metal ring, such that motion of the electrons are confined in a ring with radius $R$. We choose a reference frame in which the ring lies in the $xy$-plane and is centered on the origin, and the dynamical variable is only the 
azimuthal angle variable $\varphi$ in this plane. In presence of an universal external magnetic field $\vec{B}= (0, 0, B)$, according to the 
Hamiltonian given in \eqref{eq:nonRelNCH}, the corresponding static noncommutative 
Schr\"{o}dinger equation is
\begin{equation}\label{eq:ncsch}
\frac{1}{2m_{\theta}}
\left(-i\hbar \frac{1}{R} \frac{d\psi}{d \varphi} + \frac{eB_{NC} R }{2} 
%\left(-i\hbar \frac{1}{r} \frac{d\psi}{d \phi} + \frac{eBr}{2} Q\frac{\Phi_{NC}}{L} 
 \right)^{2}
\psi_{n}= \epsilon_{n}\psi_{n} \,,
\end{equation}
where $B_{NC} = B - B_{\xi}$ is the noncommutative corrected magnetic field, and the symmetric gauge was used in above derivation. Furthermore, we have set $Q=-e$ for electrons. The solutions of the \eqref{eq:ncsch} are $\psi_{n} =  e^{i n \varphi}/\sqrt{2\pi}$ with 
integral quantum numbers $n$, and energy levels
\bee
\epsilon_{n} = \frac{ h^{2} }{ 2m_{\theta} L^{2} } \left(n + \frac{\phi_{NC}}{\phi_{0}}\right)^{2}\,,
\ene 
where $\phi_{NC}= \pi R^{2}B_{NC}$ is the total magnetic flux and $L=2\pi R$ is the length of the metal ring. Similar to the results in ordinary phase space, the energy eigenvalues $\epsilon_{n}$ are parabola in the magnetic flux $\phi_{NC}$ for a given quantum number $n$, however it is periodic overall in magnetic flux $\phi_{NC}$ with even spacing of the different $\epsilon_{n}$.
This property results in periodic velocity and hence charged current.
By using the Heisenberg equation, one can easily obtain the expectation value of the one-particle mechanical velocity operator at the quantum state $\psi_{n}$,%$|n\rangle$,
\bee\label{eq:velocity}
v_{n} = \frac{ h }{m_{\theta}L}\left(n + \frac{\phi_{NC}}{\phi_{0}}\right)\,.
\ene
The corresponding charged current can be obtained by multiplying $v_{n}$ by $-e/L$ since the electron turns through a given point $v_{n}/L$ times in an unite time,
\bee\label{eq:currentSL}
i_{n} 
%= - \frac{e}{L} v_{n} 
= - \frac{ eh }{m_{\theta}L^{2}}\left(n + \frac{\phi_{NC}}{\phi_{0}}\right)\,.
\ene
While persistent currents are usually associated with superconductors, it can also exist in some resistive material as well, as long as the mean scattering length is considerably larger then the length of ring~\cite{PhysRevB.37.6050,Bleszynski-Jayich272,PhysRevB.81.155448}. The current \eqref{eq:currentSL} is the single-level contribution to the persistent charged current. In case of a real ring, currents from many electrons have to be summed up. In consideration of that the long-range interactions within the metal is suppressed due to the charge screening effect, we will neglect the interactions among those electrons. This approximation is not straightforward because the noncommutative correction is expected to be small. Therefore this approximation is valid only in the case of that the noncommutative correction is larger then the contributions of long-range interactions. Here we simply assumed this approximation is justified, and discuss it when the temperature effect is included. 

For $0<\phi_{NC}<\phi_{0}/2$, the energy levels are arranged in ascending order, \ie, $0$, $-1$, $+1$, $-2$, $+2$, $\ldots$. For  $-\phi_{0}/2<\phi_{NC}<0$, the ordering of levels is reversed for each $n$. Because of the linearity of the current in quantum number $n$, the contributions from positive integer $+n$ and negative integer $-n$ are cancelled with each other. Therefore in case that levels are filled by both positive and negative quantum number $n$, the net contribution is linearly proportional to the magnetic flux $\phi_{NC}$. Denoting the highest level filled with both positive and negative integers as $N$, then the total current in this case is
\begin{equation}\label{eq:currentN}
I_{N} = - \frac{eh}{m_{\theta}L^2}(4N+2)\frac{\phi_{NC}}{\phi_{0}}\,.
\end{equation}
Here, we have taken into account of the twofold degeneracy of each level due to spin. This result is also valid for $-\phi_{0}/2<\phi_{NC}<\phi_{0}/2$ since the cancellation explained above happens in spite of the sign of the magnetic flux $\phi_{NC}$. The contributions of next four electrons to the total current are,
\begin{eqnarray}
I_{N+1}^{1} 
&=& 
I_{N} - 
\frac{eh}{m_{\theta}L^2}
\left( - \sigma (N+1) + \frac{\phi_{NC}}{\phi_{0}}\right)\,,
\\[2mm]
I_{N+1}^{2} 
&=& 
I_{N} - 
\frac{2eh}{m_{\theta}L^2}
\left( - \sigma (N+1) + \frac{\phi_{NC}}{\phi_{0}}\right)\,,
\\[2mm]
I_{N+1}^{3} 
&=& 
I_{N} - 
\frac{eh}{m_{\theta}L^2}
\left( - \sigma (N+1) +3 \frac{\phi_{NC}}{\phi_{0}}\right)\,,
\\[2mm]
I_{N+1}^{4} 
&=& 
I_{N} - 
\frac{4eh}{m_{\theta}L^2}\frac{\phi_{NC}}{\phi_{0}}\,,
%\\
%I_{N+3} 
%&=& 
%I_{N} - 
%\frac{eh}{m_{NC}L^2}\left( - \sigma N - \frac{\phi_{NC}}{\phi_{0}}\right)
\end{eqnarray}
where $\sigma={\rm sgn}(\phi_{NC})$ is the sign of $\phi_{NC}$, and the superscript of $I_{N+1}$ is used to indicate the number of electrons that have been added to the system. One can clearly see that, due to the cancellation, the total current always has a magnitude of the order of the current contribution from the highest energy.
%the current generated by the electrons added to the ring tends to cancel out the contribution of the previous electron resulting in a total current with a magnitude of the order of the current contribution from the highest energy level.
At a static state with $n=N$, according to \eq{eq:velocity}, the Fermi velocity of the $N^{\rm th}$ level at $\phi_{NC}=0$ is $ v_{F} = hN/m_{\theta}L$.
Therefore, no matter how the energy levels are filled, the typical magnitude the total current is of order $I_{0} = ev_{F}/L$.
Since this value has the same magnitude as that of the electron in the highest occupied level, each
additional electron has a strong effect on the flux dependence of the current. Therefore, the magnetic flux dependence is sensitive to the number of electrons in the ring. In Ref.~\cite{Liang:2014ija}, the authors studied the currents by distinguishing even and odd number of electrons. However, for an ensemble of rings, the number of electrons has a spread, hence averaging the currents among different number of electrons is necessary. Therefore we will don't study the currents separately, we are aiming to find observables that are independent of the number of electrons.

From above results one can see that the charged current receives noncommutative corrections in two ways. The first one is through the renormalization of the electron mass, and hence of the Fermi velocity $ v_{F} = hN/m_{\theta}L$. This correction is independent of the number of electrons (in the limit $N\gg 1$).
However, the noncommutative correction is too small, for $\sqrt{\theta} \sim 1{\rm fm}$ the correction is at the order of $10^{-15}$. Therefore for an ideal ring the correction due to noncommutativity of coordinate operators is completely negligible. In case of finite temperature, the Fermi velocity is affected by thermal fluctuation, hence there is a hope to probing the $\theta$-dependent correction through the thermal behavior. However, we will shown in next section that this is also challengeable.  

The second one is violation of symmetry property of the persistent current.
In the ordinary phase space, \ie, $\phi_{NC} = \phi = \pi R^2B$, the current is antisymmetric about $\phi_{NC}=0$, \ie, $I(-\phi_{NC})= -I(\phi_{NC})$, and hence the null-point is defined by the condition $B=0$, \ie, $I(- B ) = - I(B )$. Nevertheless, on noncommutative phase space, this null-point is shifted by the noncommutative correction which is independent of the external magnetic field $\vec{B}$, \ie, $I(- B ) = - I(B ) + 2 I_{\xi}$, where $I_{\xi}$ denotes the $B$-independent part of the current $I(B)$. Therefore an asymmetric observable can be defined to probe the noncommutativity of the momentum space. Here we introduce following asymmetry observable to 
measure the noncommutative effects,
\bee\label{eq:asyDefine}
\mathcal{A}_{\xi} 
= \frac{  I(B ) + I (- B )  }{  I(B ) - I (- B )  } 
= \frac{ I_{\xi} }{ I_{B} } \,,
\ene
where $I_{B}$ denotes the $B$-dependent part of the current $I(B)$. For an idea ring, the persistent current is given by \eqref{eq:currentN}, thus the asymmetry can be written as
\bee
\mathcal{A}_{\xi} 
= - \frac{1}{e\hbar} \frac{ \xi }{ B }\,.
\ene
Clearly, this asymmetry vanishes in the limit $\xi\to 0$. On the other hand, the 
asymmetry $\mathcal{A}_{\xi} $ is independent of the noncommutative parameter $\theta$, hence it measures only the noncommutativity of the momentum space. Furthermore, the sign of the noncommutative parameter $\xi$ can also be measured. This asymmetry is also affected by thermal fluctuation, and the details are given in next section.

\section{Temperature Dependence}\label{sec:temperature}
\begin{figure}[h]
\begin{center}
\includegraphics[scale=1.]{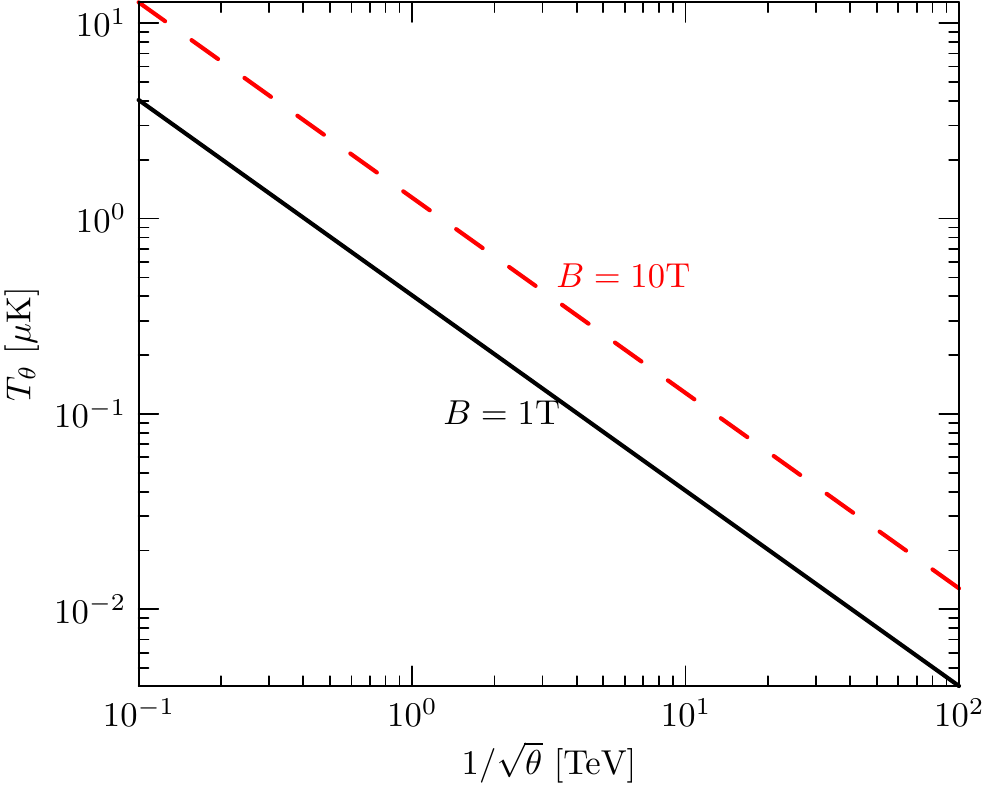}
\caption{Critical temperature $T_{\theta}$ for observing the noncommutative 
correction on the total current. The black-solid line shows $T_{\theta}$ for external 
magnetic field $B=1{\rm T}$, while the red-dashed line shows the case of 
$B=10{\rm T}$. In addition, a typical Fermi velocity 
$v_{F}=1\times10^{6}{\rm m\cdot s^{-1}}$ is used in both cases.} 
\label{fig:temperature}
\end{center}
\end{figure}

In last section we have obtained the noncommutative corrections on the persistent charged current generated by external magnetic field in a metal ring. In this section we study their temperature dependence. For convenience we will use another expression of the sing-level current to study the thermal effects. In last section, we derive the charged current by using the Heisenberg equation. In fact it can also be obtained by differentiating the energy with respect to external magnetic flux,  
\bee
i_{n} = - \frac{\partial \epsilon_{n} }{ \partial \phi }\,.
\ene
We have checked that these tow approaches lead to same result in our case. In this approach, the total current at temperature $T$ can be written as 
\bee\label{eq:currentEnergyDefine}
I = \sum \bigg(- \frac{\partial \epsilon_{n} }{ \partial \phi } \bigg) 
f( \epsilon_{n}, \mu, T )
= - \frac{ \partial \Omega}{ \partial \phi }\,,
\ene
where $f( \epsilon_{n}, \mu, T )$ is the Fermi-Dirac distribution function, 
and $\Omega$ is the grand potential of thermodynamics. In our case, 
the grand canonical potential can be written as
\begin{equation}
\Omega 
= 
-\frac{1}{\beta}\int_{-\infty}^{\infty} d\epsilon \varrho(\epsilon, \phi_{NC})
\ln\left(1+ e^{-\beta(\epsilon-\epsilon_{F})} \right)\,,
\end{equation}
where $\beta^{-1}=k_{B}T$; $\epsilon_{F}$ is the Fermi energy, 
and $\varrho(\epsilon, \phi_{NC})
= \sum_{n} g_{n}\delta(\epsilon-\epsilon_{n}(\phi_{NC}))$ 
is the flux-dependent density of states with degeneracy $g_{n}$. In the 
following analysis we will neglect the spin-dependent interactions, and in this 
approximation $g_{n}=2$. Rewriting the density of states in terms of its hamornics, and performing integration by parts, one can obtain
\begin{equation}
\Omega 
=
\int_{-\infty}^{\infty} d\epsilon
\left\{
\sum_{p>0} \frac{4\hbar^2k\cos(2\pi p \nu_{NC})}{m_{\theta} \pi p^2 L} 
\left( \frac{\sin(pkL)}{pkL} - \cos(pkL)\right)+\frac{8}{3}
\sqrt{\frac{2m_{\theta}L^2}{h^2}}\epsilon^{3/2}
\right\}
f'(\epsilon, \epsilon_{F}, T)\,,
\end{equation}
where $\nu_{NC} = \phi_{NC}/\phi_{0}$, and the function $f'(\epsilon, \epsilon_{F}, T)$ is the derivative of Fermi-Dirac
distribution, and given by following expression
\begin{equation}
f'(\epsilon, \epsilon_{F}, T)
=
\frac{1}{4}\beta\sinh^{2}\left[\frac{1}{2}\beta(\epsilon-\epsilon_{F})\right]\,,
\end{equation}
which is peaked around the Fermi energy $\epsilon_{F}$ with 
characteristic width $\beta^{-1}$.
The corresponding current can be obtained by using \eqref{eq:currentEnergyDefine},
\begin{equation}\label{eq:fullcurrent}
I=
-\int_{-\infty}^{\infty} d\epsilon
\sum_{p>0} \frac{8\hbar^2k \sin(2\pi p \nu_{NC}) }{m_{\theta} p L\phi_{0} }
\left( \frac{\sin(pkL)}{pkL} - \cos(pkL)\right)
f'(\epsilon, \epsilon_{F}, T) \,.
\end{equation}
Typically, the temperature $k_{B}T$ of the metal system is much less then the 
Fermi energy $\epsilon_{F}$, \ie, $k_{B}T \ll \epsilon_{F}$, therefore, the 
factor $k$ will be of order $k_{F}=\sqrt{2m_{\theta}\epsilon_{F}}/\hbar$ in the 
integral over $\epsilon$. For typical Fermi velocity $v_{F} \approx 10^{6}{\rm m/s}$, $k_{F} \approx 8.7\times10^{9}{\rm m^{-1}}$, the product 
$k_{F}L = 8.7 \times 10^{3}$ is much larger then $1$ for a typical length 
of ring $L=1\mu{\rm m}$. Therefore, contributions from the oscillation term 
$\sin(pkL)/(pkL)$ in the integral \eqref{eq:fullcurrent} can be safely neglected. 
The contributions from the term proportional to $\cos(pkL)$ can be calculated
perturbatively by expanding $k$ around $k_{F}$,
\begin{equation}
k\approx k_{F}(1+0.5r + \mathcal{O}(r^2) ) \,,
\end{equation}
where $r= \epsilon/\epsilon_{F} -1$ is a small quantity as explained above. In 
the following calculations, we will keep only the linear terms of $r$. This approximation simplifies the calculations, however some conditions must be satisfied in order to 
get reasonable noncommutative corrections. The leading order noncommutative correction on $k$ is given by
\bee
\hbar \delta_{NC} k = \frac{ \theta B }{ \phi_{0} } mv_{F}\,,
\ene
while the second order thermal contribution is given by
\bee
\hbar \delta_{th.} k = \frac{1}{8} \bigg( \frac{ k_{B} T }{ \epsilon_{F} } \bigg)^{2}mv_{F}\,.
\ene
Therefore the critical temperature, above which the $\theta$-dependent noncommutative correction will be washed out by the thermal motion, is given as follows
\bee
T_{\theta} = \frac{2m v_{F}^{2}}{k_{B}} \sqrt{ \frac{\theta B}{ \phi_{0} }} \,.
\ene
In Fig.~\ref{fig:temperature} we shows the dependence of the critical 
temperature $T_{\theta}$ on noncommutative parameter $\theta$. The black-solid 
line shows $T_{\theta}$ for external magnetic field $B=1{\rm T}$, while the red-
dashed line shows the case of $B=10{\rm T}$. Since the Fermi velocity 
$v_{F}$ is relatively stable among most of metal, we have used a typical 
value $v_{F}=1\times10^{6}{\rm m\cdot s^{-1}}$ in the estimation. We can see
from the figure that for a space noncommutativity at the ${\rm TeV}$ scale, 
a temperature bellow $1\mu{\rm K}$ is necessary, which is challengeable for now. Therefore in the following discussions, we will neglect this correction.

\begin{figure}[t]
\begin{center}
\includegraphics[scale=1.]{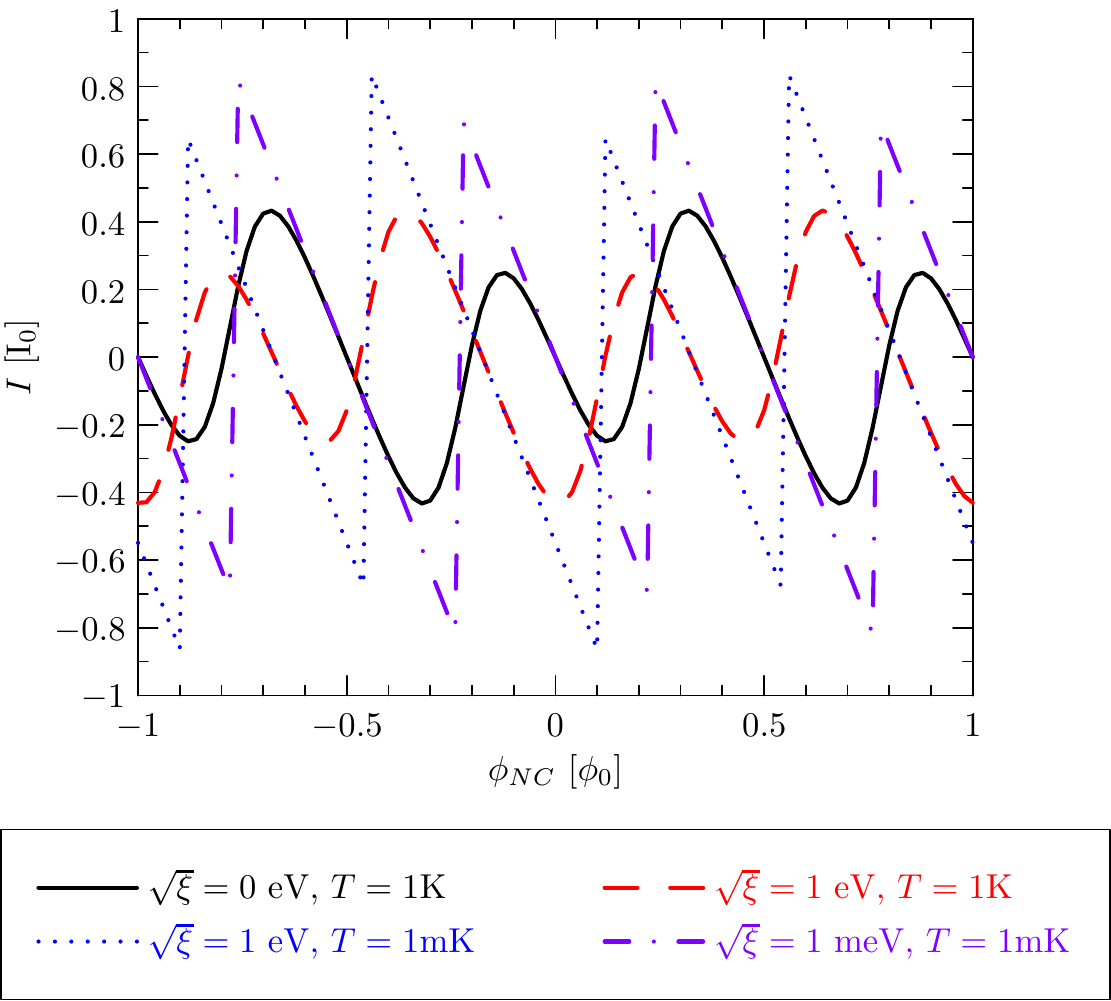}
\caption{Oscillation and the shift of the null-point of the total current in case of finite temperature. The total current is given in unite of $I_{0}$ and the magnetic flux $\phi_{NC}$ is given in unite of $\phi_{0}$. The black-solid line shows the total current for $\xi=0$ and $T=1{\rm K}$. The red-dashed line shows the case for $\sqrt{\xi}=1{\rm K}$ and $T=1{\rm K}$. The blue-dotted line shows $I$ for $\sqrt{\xi}=1{\rm eV}$, $T=0.1{\rm mK}$, and purple-dash-dotted line is for the case of $\sqrt{\xi}=1{\rm meV}$, $T=1{\rm mK}$. In all cases, 
$v_{F}=1\times10^{6}{\rm m\cdot s^{-1}}$ and $L=1{\rm \mu m}$ are used.} 
\label{fig:current}
\end{center}
\end{figure}

In this approximation the temperature-dependent current is given as
\begin{equation}\label{eq:finiteTemCurrent}
I 
= \frac{4}{\pi}I_{0} 
\sum_{p>0} p^{-1} \sin(2\pi p \, \nu_{NC}) \cos(pk_{F}L) 
g(\kappa_{p})\,,
 %\frac{ \kappa }{ \sinh(\kappa) }
\end{equation}
where the factor $\kappa_{p}= T/T_{p}$ with a typical temperature $T_{p}=\hbar v_{F}/(2\pi p k_{B} L)$, and the function $g(\kappa_{p})$ is defined as
%the typical temperature $T_{p}$ is
%\begin{equation}
%T_{P} = \frac{1}{\pi p k_{B}} \frac{\hbar^{2}}{2m_{\theta}}\frac{k_{F}}{L}
%\end{equation}
\begin{equation}
g(\kappa_{p})=\frac{ \kappa_{p} }{ \sinh(\kappa_{p}) }\,.
\end{equation}
For $T>T_{p}$ , the $p^{\rm th}$ harmonic of the current decays exponentially 
as temperature increasing. Furthermore, the total current is sinusoidal in the magnetic flux, which is different from the linearity as given in \eqref{eq:currentN}. Fig.~\ref{fig:current} shows such oscillations in cases of different temperature and noncommutative parameters $\xi$. The black-solid line shows the total current at $T=1{\rm K}$ on ordinary phase space. As expected, it pass through the null-point.
The red-dashed line shows the total current at the same temperature but $\sqrt{\xi}=1{\rm eV}$. One can see clearly that the null point has been shifted. The blue-dotted line shows the total current with $\sqrt{\xi}=1{\rm eV}$, $T=1{\rm mK}$. As expected, the magnitude of the total current is enhanced by lowering the temperature. The purple-dash-dotted line shows the case of $\sqrt{\xi}=1{\rm meV}$, $T=1{\rm mK}$. One can see that an approximately linear dependence is recovered, and the null-point closes to the origin. 

\begin{figure}[t]
\begin{center}
\includegraphics[scale=1.]{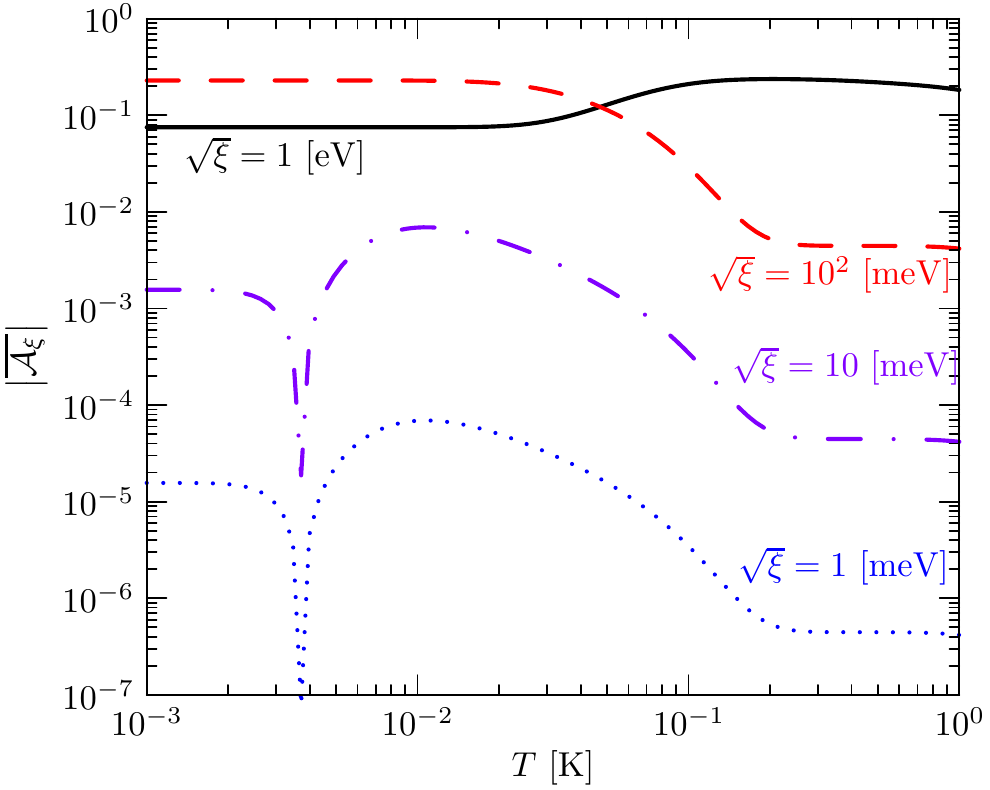}
\caption{Temperature dependence of the magnitude of the averaged asymmetry. The black-solid, red-dashed, purple-dash-dotted, blue-dotted lines show the cases of $\sqrt{\xi} = 1{\rm eV}$, $\sqrt{\xi} = 10^{2}{\rm meV}$, $\sqrt{\xi} = 10{\rm meV}$, $\sqrt{\xi} = 1{\rm meV}$, respectively. In all cases, 
$v_{F}=1\times10^{6}{\rm m\cdot s^{-1}}$ and $L=1{\rm \mu m}$ are used.} 
\label{fig:asymmetry}
\end{center}
\end{figure}

Furthermore, according to the definition of the asymmetry given in \eqref{eq:asyDefine}, the temperature dependent asymmetry is
\bee
\mathcal{A}_{\xi} 
=-
\frac{ \sum_{p>0} p^{-1} \sin( 2\pi p \nu_{\xi} ) \cos(2\pi p \nu) \cos(pk_{F}L) g(\kappa_{p}) }
{ \sum_{p>0} p^{-1} \cos( 2\pi p \nu_{\xi} )\sin(2\pi p \nu) \cos(pk_{F}L) g(\kappa_{p}) }\,,
\ene
where $\nu=\phi/\phi_{0}$ and $\nu_{\xi}=\phi_{\xi}/\phi_{0}$.
However, since denominator of $\mathcal{A}_{\xi}$ is sinusoidal in the filling factor $\nu$, and hence occasionally the value of $\mathcal{A}_{\xi}$ can be unreasonably huge. Therefore, it is more convenient to define an averaged asymmetry over a certain range of the external magnetic flux. In consideration of that the terms with $p=1$ have relatively larger contributions, we chose the range $0 \le \nu \le 1/4$ as a benchmark point, and the corresponding averaged asymmetry is given as
\bee
\overline{\mathcal{A}_{\xi} }
=-
\frac{ \int_{0}^{1/4} d \nu  \sum_{p>0} p^{-1} \sin( 2\pi p \nu_{\xi} ) \cos(2\pi p \nu) \cos(pk_{F}L) g(\kappa_{p}) }
{  \int_{0}^{1/4} d\nu  \sum_{p>0} p^{-1} \cos( 2\pi p \nu_{\xi} )\sin(2\pi p \nu) \cos(pk_{F}L) g(\kappa_{p}) }\,.
\ene
Fig.~\ref{fig:asymmetry} shows the temperature dependence of the magnitude of the averaged asymmetry for different values of the noncommutative parameter $\xi$. Oscillation of the asymmetry is due to that the relative importance of the contribution of the $p^{\rm th}$ harmonic is affected by the temperature.
From the figure one can also see that for $\sqrt{\xi} \sim {\rm meV}$ the asymmetry $\overline{\mathcal{A}_{\xi} }$ is at an order of $10^{-4}$.
%$\big|\overline{\mathcal{A}_{\xi} }\big|$. 

\section{Conclusions}\label{conclusion}
In summary, we have studied the noncommutative corrections on the persistent charged current in a metal ring, as well as their temperature dependence. Based on the SW map, the non-relativistic Hamiltonian on a noncommutative phase space was obtained by using the FW transformation. At the leading order of noncommutative parameters, the corrections due to coordinate and momentum noncommutativity come into play through the mass renormalization and shift of the external magnetic flux, respectively. Due to that the mass renormalization affects only the total magnitude of the current, it is challenging to measure the coordinate noncommutative parameter $\theta$. However, the noncommutativity of momentum can violate symmetry property of the persistent current. Furthermore, this violation effect still exist at a finite temperature. Based on this violation effect, we have introduced an asymmetry observable, which is sensitive to the momentum noncommutativity, to measure the corresponding parameter $\xi$. In practice, the finite cross section of the metal ring affects also the persistent current. However, the symmetry property is preserved even in this case. Therefore, the observable introduced in our paper is essential to probe the momentum noncommutativity.

\noindent\textbf{Acknowledgments}: 
K.M. is supported by the China Scholarship Council, and the National Natural Science Foundation of China under Grant No. 11647018, and partially by the Project of Science and Technology Department of Shaanxi Province under Grant No. 15JK1150.

\bibliography{aString}

\end{document}